\documentclass[12pt]{article}

\usepackage{graphicx}
\usepackage{amssymb,amsmath}
\begin{document}

\title{Modelling of radiation defects evolution in regular structures by the cellular automata method}
\author{K. S. Baktybekov, S. G. Karstina, E. N. Vertyagina\\ Karaganda State University named after E. A. Buketov, Kazakhstan}
\maketitle
\begin{abstract}
The results of modeling of radiation defects formation and evolution on the surface and in the volume of a crystal are presented in this article. Statistical properties are calculated for the investigated system. It is revealed that defects structure is a multifractal and system entropy decreases, while observing self-organization of the physical system.


\end{abstract}
\section{Introduction}
\label{}
A cellular automata (CA) is widely used for modeling of the systems, in which the spatial interaction between the elements has an important role. CA provides free choice of the structure type and rules of the system development. It allows simulating on its base and solving a range of problems. Moreover, any physical system that can be described by differential equations, can be approximated as CA by using the finite differences and discrete variables.

Method of modeling of radiation defect formation in solids by making use of CA is offered in this paper. Defects distribution on surface and in volume of solids and defects evolution is the fundamental scientific interest, since this is an example of self-organized system. Method of multifractal analysis is used for determination of statistical parameters of investigation system. Further researches will allow predicting radiation stability of solids.

 \section{Cellular automata method}
 \label{}
CA is a universal model of parallel calculations. This is a discrete dynamic system, which represents a set of equal nodes or cells combined identically. Nodes form the lattice of the cellular automata. Lattices can be of any type, differing both on dimension and on the cells form. Each node state is completely defined by the nearest nodes states and its own state. All nodes develop simultaneously according to the identical rules.

The extensive experimental study of CA, performed by Wolfram \cite{Wolfram}, says that all models, generated in the CA evolution process from the disordered initial state, can be divided into four classes. CA, which belongs to the IV class, are the most interesting for modeling natural phenomena. In most cases all nodes reach zero value in a finite time interval. However, in some cases propagating configurations or stable periodic structures are formed according to the local interaction rules. Those can be stable infinitely.  It is possible to simulate processes of radiation defects formation in solids and defects evolution by making use of the IV class cellular automata.
\section{Method of modelling of the radiation defects formation processes in solids}
 \label{}
The choice of CA as a method of modeling is explained by the fact that solid state consists of a large number of elementary units. Each unit has a simplest structure and performs the simple function. However, when they are combined together by local interactions, the result, as a rule, is a complex behavior of the system as a whole. At the microscopic level CA nodes can represent atoms of the crystal lattice.
Irradiated solid state with face-centered cubic crystal lattice is chosen as a model of the research. Electronic excitation is formed in crystal under ionizing radiation. Their nonradiative disintegrations can lead to formation of Frenkel or other defects pairs. Irradiation source generates defects in solid state continuously. Simultaneously the recombination processes occur in the solid.

The considered system is open and nonlinear. In the systems, which are under external action of the energy or particles stream, comparatively stable states, which are not corresponding to a minimum of free energy, can be formed. Temporal or spatial orderliness, self-organization is often observed for these states (called dissipative structures).
The CA rules for the systems can be defined correctly if they satisfy the sequences of the conditions and they are invariable at symmetry transformation such as rotation and mirror reflection. The indicated model completely satisfies the CA use condition since possesses the following characteristics:

1) cubic face-centered crystal lattice is homogeneous, defect formation occurs in the nearest nodes of anionic sublattice accidentally;

2) nodes states set takes three values: -1, 0 and 1, where 0 corresponds to the defectless node,  1 and 1 correspond to the conditionally negative and conditionally positive defect respectively;

3) the minimum distance between anions is chosen as the radius of defects interaction. Thereby, the rule of local interactions is also satisfied. In a flat endless lattice each node interacts with four neighbors, in a three-dimensional lattice number of neighbors is 12.

The following input data are required for using the model:

1) plane edge size (for modeling on a surface), or cube edge size (for modeling in a volume);

2) irradiation dose power (is defined by initial density of accidentally distributed defects in investigated system);

3) irradiation duration of the system (is assigned by amount of cycles of the modeling program blocks);

4) point number for calculations (modeling irradiation process is stopped at the point, the snapshot of the current defect location in the system is created, the most important statistical parameters are calculated).

Two blocks of the modeling program ensure the condition of parallel changing the nodes states in lattice. The first block realizes formation of defects, the second - probabilistic annihilation of conditionally inverse charged defects.
\section{Results and discussion}
 \label{}
The described model was used for calculation on a flat lattice with the edge size of 500 nodes ($25 \cdot 10^4$ elements) and in a cubic lattice with the edge size of 100 nodes ($10^6$ elements). Irradiation duration of the surface corresponded to 500 thousands, volume - 50 thousands cycles at dose power, equivalent to 0.1\% initial defects concentration.

Spatial division of inverse charged defects and formation of one type charged defects aggregation is observed in system at long irradiation simulation. Concentration of radiation-induced defects in accumulations is more, than their average concentration in the whole crystal. Aggregating centers formation within accumulations is more probably, than in the case of homogeneous distribution of defects in a crystal volume. Creation of the ordered structure from radiation defect conglomeration under radiation is a synergetic effect, which indicates self-organization of the structure, was found in chaotic state before.

The simplest statistical parameter characterizing CA configuration is the average density of the nonzero nodes. General density of accumulated defects increases in several times in modeling system and than remains constant. The speed of saturation is as high as irradiation dose in identical size solids.

Irreversible nature of evolution from the disordered initial configuration leads to the self-organization phenomena, generation of self-similar models with fractal dimension. In the evolution process of the model in two- and three-dimensional lattices it was revealed that defect structure represents dynamic fractal of the disordered configuration.

Calculation of information entropy as a measure of disorder was performed for the system. Entropy of a fractal structure can be considered as a logarithm of the average number of the possible system states:
\begin{align}\label{eq:entropy}
S &= -\sum_{i}p_i \ln{p_i},&  \sum_{i}p_i &= 1,
\end{align}
where $p_i$  is probability to find a certain node of the lattice in the state numbered $i$.

\begin{figure}
\begin{center}
\includegraphics*{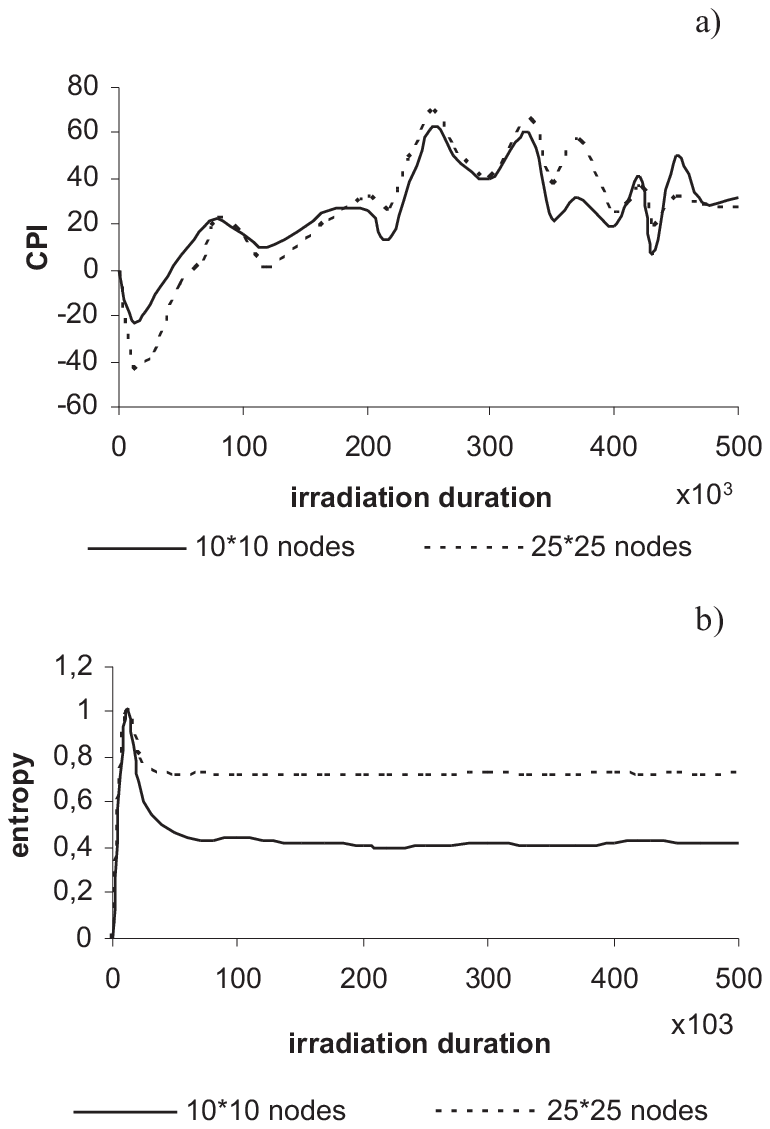}
\end{center}
\caption{Dependencies of the conditional potential of the interaction CIP (a) and normalized specific entropy (b) on irradiation duration of 500*500 nodes surface for two sizes of conglomerations, which are present in system.}
\label{f1}
\end{figure}

Fig.\ref{f1}a shows that calculated for the modeling system normalized specific entropy \cite{Zhanabaev}, decreases and reaches a constant value that differs for various aggregation sizes. The defect density in the system, as it was mentioned above, is also remains constant even if defects configuration is changed in the lattice all the time of evolution. This indicates that there are a process of self-organization of defects structure in crystal and stability of the formed system.

To estimate the typical size of the formed aggregations we introduced a statistical parameter, named "conditional potential of the survived defects interaction" (CPI) \cite{cpi}. This value is calculated for nonzero nodes pairs according to the formula
\begin{equation}\label{eq:cpi}
U = \frac{1}{2} \sum_{k} \sum_{i,j} \frac{q_i q_j}{r_{ij}},
\end{equation}
where $i, j$ - indexes of pair components, $k$ - an value of interaction nodes radius, at which it is possible to split the lattice, $q_i, q_j$ - nonzero nodes (the equally charged defects), $r_{ij}$ - the distance between interacting nodes.

The CPI dependence on irradiation duration allows watching the evolution of the defects structure distribution in solids (see fig.\ref{f1}b). However, the aggregation curves change identically at different typical size of defects accumulations. This again reveals ordering process taking place in the system and fractal structure of formed aggregation. Kinetic curves have quasi-oscillating shape at stage of saturation. It can be explained by dynamic processes of growth of formed clusters and their destruction.

It is noticeable that the CPI minimum falls at the information entropy maximum, corresponding to chaotic distribution of defects on a surface and in a volume. Decrease of the system entropy and growth of CPI indicate observed effect of self-organization of defects structure.

For quantitative estimate of the structured conversions, occurring in solids under irradiation, snapshots of current defect location were analyzed at different time moments. The snapshots were covered by a net with changing cell size. Let the cells edge size is marked as  $l$, and amount of cells, used for covering of fractal is marked as  $N(l)$. Hence the fractal dimension of forming defect structure can be calculated via the following methods.

1) Hausdorff dimension:
\begin{equation}\label{eq:dimens}
D = \frac{\ln{N(l)}}{\ln{l}}
\end{equation}

\begin{figure}
\begin{center}
\includegraphics*{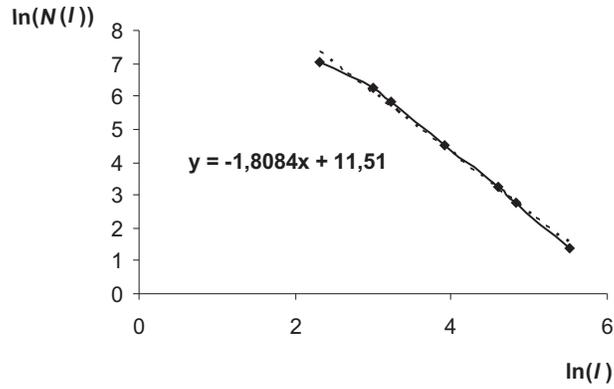}
\end{center}
\caption{The graphic method for determination of dimension. The approximating straight line is marked by dotted line.}
\label{f2}
\end{figure}

2) Graphic method for determination of dimension. Dependence  $N(l)$ on $l$ is built in double logarithmic coordinates (see fig.\ref{f2}). Taken with inverse sign argument coefficient defines the fractal dimension in equation of approximating straight line. \\
Both methods give a crude estimation of dimension: from 1.59 to 1.96 for a plane, or from 2.64 to 3.0 for a cube. These values do not contain information about statistical properties of aggregates.

The data of snapshots allows considering formed aggregate as multifractal, because self-similarity is not exactly realized in real physical systems. The degree of aggregate regularity is determined for a multifractal set, which consists of several subsets with different fractal dimension \cite{mfa}.

\begin{figure}
\begin{center}
\includegraphics*{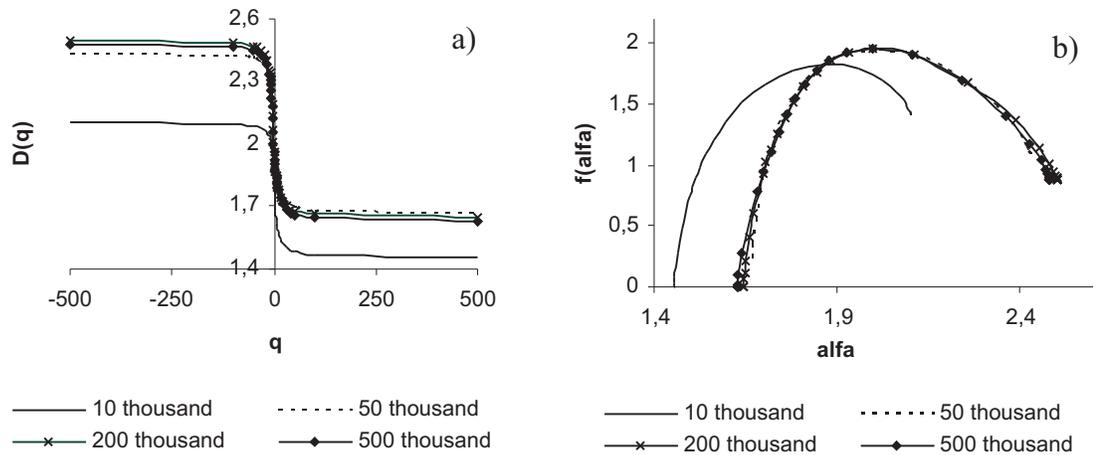}
\end{center}
\caption{Change of multifractal parameters of the defects structure with size of 25*25 nodes at irradiation of the planes with size of 500*500 nodes at different time moments: a) Renyi generalized dimensions spectrum; b) multifractal spectrum function.}
\label{f3}
\end{figure}

The charts are constructed for the generalized dimensions $D(q)$ and the value of the self-organizing control parameter $q$, and for the multifractal spectrum function $f(\alpha)$ (see fig.\ref{f3}). The set of different values of function $f(\alpha$) at distinguish $\alpha$ is a spectrum of fractal dimensions for homogeneous subsets, by which the initial set is divided. As it was calculated, $f(\alpha$) changes within 0.24 to 1.96 on a plane or within 0.8 to 2.68 in a volume.

The method of multifractal analysis is capable to distinguish similar subsystems and recognize the detail of the structure, as well as allows getting information about stability of the structure as a whole and revealing the degree of self-organization of formed structures. In weakly non-homogeneous systems the $f(\alpha$) curve is well approximated by a parabola. The coefficient of reliability of square approximation $R^2$ is 0.957 for curve (see fig.\ref{f3}b), corresponding to modeling duration 10 thousand cycles, when system is corresponding to modeling duration 10 thousands cycles, when system is found in the chaotic state. At  the time moment, corresponding to 500 thousand cycles value $R^2$ decreases to 0.856 that indicates an increase of the system non-homogeneity.
\section{Conclusions}
\label{}
It is shown that defects structure evolution from the initial chaotic state leads to the synergetic effect of clusterizations in submitted paper. Continuous evolution of defects and gradual transition to stable formations is observed under irradiation. It is characterized by reduction of the system entropy. The multifractal analysis of the defects structure has shown that investigated structure is non-homogeneous and consists of several subsets, each of those is characterized by own fractal dimension.

\end{document}